\documentclass{article}

\begin{document}

\title{New Facts From the First Galaxy Distance Estimates}
\author{by Ian Steer, NED-D (isteer@hotmail.com)}
\date{Accepted for publication in JRASC, February, 2011}
\maketitle

\section{Introduction}

A new database from the NASA/IPAC Extragalactic Database (NED) of galaxy Distances (NED-D), normally the source for the newest precision-based estimates, provides access to the oldest redshift-independent extragalactic distances in the publication record. Based on review of the astronomical literature in NASA's Astrophysics Data System (ADS), we found 290 legacy distances to 62 galaxies from 37 astronomical references that were published up to and including 1930.

Two new surprises emerge when the early distance estimates are placed in chronological order. First, before Hubble gained fame by publishing 71 distances to 32 galaxies, a total of 97 distances to 48 galaxies had already been published (Hubble 1926). Second, Hubble was not the first to discover but rather first to prove, based on his distance estimates, that a universe of galaxies exists beyond the Milky Way and that this universe of galaxies is apparently expanding (Hubble 1926, 1929). Though credit is given to Hubble, these discoveries, were first made but alas not proven, by Lundmark 
(1924a, 1925).

\section{Standard Candles, Standard Rulers}

A discovery by Miss Henrietta Leavitt led to the first standard-candle method for estimating galaxy distances (Leavitt 1908, 1912). Yet the discoverer of the Cepheid period-luminosity relation remained a footnote until very recently, when Leavitt's Law of Cepheids was fully credited (Freedman \emph{et al.} 2009, Madore \emph{et al.} 2009a, 2009b, and Marengo \emph{et al.} 2010). Previously, most credit for the discovery had gone to Hertzsprung and, independently, Russell, as they were first to calibrate Leavitt's method and first to publish distances based on the method (Hertzsprung 1913, Russell 1913).

Improvements by Shapley (1918) led to a Cepheids-based distance to the Small Magellanic Cloud that remained fiduciary for decades. Hubble's first Cepheids distances werenÕt published until 1925. They were by far the best however, based on observations with the world's then-largest telescope, the 100-inch Hooker on Mount Wilson, CA, which began operating in 1918 (Hubble 1925a, 1925b).

Potentially, the first standard-candle extragalactic distance could be credited to Very based on novae (Very 1911). However, this estimate was un-calibrated and, giving less than a thousandth of today's distance to the Andromeda Galaxy (M31), unrealistic. The first calibrated novae-based distance, albeit to no particular galaxy but rather Òdistant spirals,Ó was made by Curtis (1917), revealing an average distance of 6 Megaparsecs (Mpc) to a collection of galaxies, and changing the culture of astronomical thinking to accept much larger extragalactic distances than had been the earlier practice. Lundmark provided the first calibrated novae distances to a specific galaxy, publishing two estimates for the Andromeda Galaxy (Lundmark 1919). One year later, Curtis (1920) used novae to determine two new distances for M31, one of 0.23 Mpc and a second of 1.2 Mpc. The average of Curtis's distances placed Andromeda at 0.7 Mpc, within 7 percent the modern value of $0.75 \pm 0.02$ Mpc (Freedman \emph{et al.} 2001) derived by NASA's Hubble Space Telescope (HST) Key Project (KP).

A third standard candle method used today - by brightest stars - was first employed by Shapley (1917) to place M31 at 0.31 Mpc. Others followed, including Lindemann \& Lindemann (1919), Shapley (1923), and Lundmark (1921, 1924b, 1925), but it was Hubble (1926) who first used a Cepheids-calibrated version of the brightest stars method to provide accurate distances for 32 individual galaxies.

The first globular cluster standard-ruler measurement using average cluster radii was made by Shapley (1922), who used the method to place the Large Magellanic Cloud at 35 kpc, within 27 percent of the latest multi-methods-based estimate of 48 kpc by Freedman \& Madore (2010), a value believed precise to within 3 percent. The technique is still in use today, particularly in NASA's HST Advanced Camera for Surveys (ACS) Virgo Cluster Survey (Jordan \emph{et al.} 2005) and the ACS Fornax Cluster Survey (Masters \emph{et al.} 2010).

\section{Secondary Methods, Hubble and Lundmark}

Two years before publication of Hubble's 1926 paper showing that galaxies were at very large distances, Lundmark used his nova distance of 0.2 Mpc to Andromeda from his 1919 paper to calibrate diameter distances to 44 other galaxies (Lundmark 1924a). Legitimately, he was the discoverer that these ``nebulae'' existed as ``island universes'' at vast distances, out to his furthest, NGC 1700, at 42 Mpc. Alas, extrapolating from one nova distance based on a single untried method (diameters) to other galaxies, did not convince his research colleagues. Hubble succeeded by employing two independent methods, both calibrated by Cepheids, one based on brightest stars and another based on galaxy apparent magnitudes to show conclusively that the distances indicated were reliable.

Lundmark (1925) actually hit upon Hubble's Law years before Hubble, writing ``more distant spirals have higher space-velocity.''  It was 
however Hubble (1929) who put the distance-velocity relationship on a quantitative footing, justifying the future use of the term ``Hubble's Law.''

\section{Others}

The first extragalactic distance estimate was published by Nichol (1840), who determined a value of 0.6 Mpc for nearby spirals in general - not bad compared to, for instance, Andromeda at 0.75 Mpc. Opik (1922) estimated the distance to Andromeda at 0.45 Mpc, within 40 percent of the HST KP value, using an early Tully-Fisher relation based on spiral rotational velocity and galaxy brightness. Finally, a dozen improbable underestimates were published, some based on apparent (but erroneous) internal rotation and others based on proper motion, impossible then, barely possible now, and now only with low precision (Vieira \emph{et al.} 2010 and references therein). Such underestimates fuelled the great debate of 1920 between Curtis and Shapley on island universes and kept the dialogue going until Hubble's 1926 paper, after which all such underestimates vanished from the record.

\section{Summary}

Credit for Hubble's discoveries of true extragalactic distances parallels that of the promotion of the heliocentric Solar System by Copernicus in 1543, the discovery of penicillin by Fleming in 1928, and of the route to North America by Columbus in 1492. Others had made earlier and similar claims Ð Eratosthenes, c. 250 BC for the Solar System, Tyndall in 1875 for penicillin, and Ericson, c. 1000 for the New World. Success went not to the first to make discoveries to their own satisfaction but rather to the first to prove them to the satisfaction of others. When it comes to the priority for discovery, 
convincing others is key.

For further reading, see Miss Leavitt's Stars, by George Johnson (W. W. Norton \& Company, 2005); The Day We Found the Universe, by Marcia Bartusiak (Random House of Canada, 2010), Man Discovers the Galaxies, by Richard Berendzen, Richard Hart, and Daniel Seeley, (Columbia University Press, 1984) and The Expanding Universe, by Robert Smith (Cambridge University Press, 1982).

\section*{Acknowledgements}

The author thanks the Editor-in-Chief for the opportunity to publish this article and for helpful suggestions, and is grateful to members 
of the NED-D Team, including Harold Corwin, for both expertise and encouragement, and gratefully acknowledges generous support from the 
California Institute of Technology and the Carnegie Institution of Canada.

\noindent
NED and NED-D staff will be at the AAS Meeting in Seattle, WA, 2011, Jan. 9-13.

\section*{References}

\noindent
Curtis, H. D. (1917). Novae in the Spiral Nebulae and the Island Universe Theory. \emph{PASP, 29,} 206.

\noindent
Curtis, H. D. (1920). Modern Theories of the Spiral Nebulae. \emph{JRASC, 14,} 317.

\noindent
Freedman, W. L. et al. (2001). Final Results from the Hubble Space Telescope Key Project to Measure the Hubble Constant. \emph{AJ, 553,} 47.

\noindent
Freedman, W. L., Rigby, J., Madore, B. F., Persson, S. E., Sturch, L. \& Mager, V. (2009). The Cepheid Period-Luminosity Relation (The 
Leavitt Law) at Mid-Infrared Wavelengths. IV. Cepheids in IC 1613. 
\emph{ApJ, 695,} 996.

\noindent
Freedman, W. L. \& Madore, B. F. (2010). The Hubble Constant. \emph{Ann.\ Rev.\ Astron.\ Astrophys., 48,} 673.

\noindent
Hertzsprung, E. (1913). \"{U}ber die r\"{a}umliche Verteilung der Ver\"{a}nderlichen vom $\delta$ Cephei-Typus. \emph{Astron.\ Nach., 196,} 201.

\noindent
Hubble, E. P. (1925a). NGC 6822, a remote stellar system. \emph{ApJ, 62,} 409.

\noindent
Hubble, E. P. (1925b). Cepheids in spiral nebulae. \emph{Obs., 48,} 139.

\noindent
Hubble, E. P. (1926). Extragalactic nebulae. \emph{ApJ, 64,} 321.

\noindent
Hubble, E. P. (1929). A Relation between Distance and Radial Velocity among Extra-Galactic Nebulae. \emph{Proc.\ Nat.\ Acad.\ Sci., 15,} 168.

\noindent
Jordan, A. et al. (2005). The ACS Virgo Cluster Survey. X. Half-Light Radii of Globular Clusters in Early-Type Galaxies: Environmental 
Dependencies and a Standard Ruler for Distance Estimation. \emph{ApJ, 634,} 1002.

\noindent
Leavitt, H. S. (1908). 1777 variables in the Magellanic Clouds. \emph{Ann.\ Harvard Coll.\ Obs., 60,} 87

\noindent
Leavitt, H. S. \& Pickering, E. C. (1912). Periods of 25 Variable Stars in the Small Magellanic Cloud. \emph{Harvard Coll.\ Obs.\ Circ., 173,} 1.

\noindent
Lindemann, A. F. \& Lindemann, F. A. (1919). Preliminary Note on the Application of Photoelectric Photometry to Astronomy. \emph{MNRAS, 79,} 343.

\noindent
Lundmark, K. (1919). Die Stellung der kugelf\"{o}rmigen Sternhaufen und Spiralnebel zu unserem Sternsystem. \emph{Astron.\ Nach., 209,} 369.

\noindent
Lundmark, K. (1921). The Spiral Nebula Messier 33. \emph{PASP, 33,} 324.

\noindent
Lundmark, K. (1924a). The determination of the curvature of space-time in de Sitter's world. \emph{MNRAS, 84,} 747.

\noindent
Lundmark, K. (1924b). The distance of the Large Magellanic Cloud. \emph{Obs., 47,} 276.

\noindent
Lundmark, K. (1925). Nebulae, the motions and the distances of spiral. \emph{MNRAS, 85,} 865.

\noindent
Madore, B. F., Rigby, J., Freedman, W. L., Persson, S. E., Sturch, L. \& Mager, V. (2009a). The Cepheid Period-Luminosity Relation (The 
Leavitt Law) at Mid-Infrared Wavelengths. III. Cepheids in NGC 6822. \emph{ApJ, 693,} 936.

\noindent
Madore, B. F., Freedman, W. L., Rigby, J., Persson, S. E., Sturch, L. \& Mager, V. (2009b). The Cepheid Period-Luminosity Relation (The 
Leavitt Law) at Mid-Infrared Wavelengths. II. Second-Epoch LMC Data. \emph{ApJ, 695,} 988.

\noindent
Marengo, M., Evans, N. R., Barmby, P., Bono, G., Welch, D. L. \& Romaniello, M. (2010). Galactic Cepheids with Spitzer. I. Leavitt Law 
and Colors. \emph{ApJ, 709,} 120.

\noindent
Masters, K. \emph{et al.} (2010). The Advanced Camera for Surveys Fornax Cluster Survey. VII. Half-light Radii of Globular Clusters in Early-type 
Galaxies. \emph{ApJ, 715,} 1419.

\noindent
Nichol, J. P. (1840). \emph{Views of the architecture of the heavens. In a series of letters to a lady.} New York, NY: H. A. Chapin \& Co.

\noindent
Opik, E. (1922). An estimate of the distance of the Andromeda Nebula. \emph{ApJ, 55,} 406.

\noindent
Russell, H. N. (1913). Notes on the Real Brightness of Variable Stars. \emph{Science, 37,} 651.

\noindent
Shapley, H. (1917). Note on the Magnitudes of Novae in Spiral Nebulae. \emph{PASP, 29,} 213.

\noindent
Shapley, H. (1918). Studies based on the colors and magnitudes in stellar clusters. VI. On the determination of the distances of 
globular clusters. \emph{ApJ, 48,} 89.

\noindent
Shapley, H. (1922). Approximate Distance and dimensions of Large Magellanic Cloud. \emph{Harvard Coll.\ Obs.\ Bul., 775,} 1.

\noindent
Shapley, H. (1923). Distance of N. G. C. 6822. \emph{Harvard Coll.\ Obs.\  Bul., 796,} 1.

\noindent
Very, F. W. (1911). Are the White Nebulae Galaxies? \emph{Astron.\ Nach., 189,} 441.

\noindent
Vieira, K. \emph{et al.} (2010). Proper-motion Study of the Magellanic Clouds Using SPM Material. \emph{AJ, 140,} 1934.

\end{document}